\magnification=\magstep1
\overfullrule=0pt
\parskip=6pt
\baselineskip=15pt
\headline={\ifnum\pageno>1 \hss \number\pageno\ \hss \else\hfill \fi}
\pageno=1
\nopagenumbers
\hbadness=1000000
\vbadness=1000000

\input epsf

\vskip 25mm
\vskip 25mm
\vskip 25mm

\centerline{\bf A BRIEF NOTE ON THE DEFINITION OF SIGNATURE } \vskip
10mm

\centerline{\bf Hasan R. Karadayi , Meltem Gungormez}

\vskip 15mm

\centerline{Dept. Physics, Fac.
Science, Istanbul Tech. Univ.} \centerline{34469, Maslak, Istanbul,
Turkey } \centerline{e-mail: karadayi@itu.edu.tr , gungorm@itu.edu.tr}

\vskip 25mm

\centerline{\bf{Abstract}}

It is known that signature of a Weyl group element is defined
in terms of the number of its simple Weyl reflections. Actual
calculations hence are not always possible especially for Weyl
groups with higher order like $E_8$ Weyl group. By extending the concept from
signature of a Weyl reflection to signature of a weight, we show that signature
of a weight is defined without referring to Weyl reflections, Though both have
the same result, the signature of a weight can be calculated for any Lie algebra.

\hfill\eject

\vskip 3mm
\noindent {\bf{I.\ INTRODUCTION }}
\vskip 3mm

Let $W({G_N})$ be the Weyl group of a Lie Algebra $G_N$ of rank N.
For $i=1,2 \dots ,N$, let $\lambda_i$'s and $\alpha_i$'s be
respectively its fundamental dominant weigths and simple roots. For
general terms on Lie Algebra technology, we refer to the excellent
book of Humphreys {\bf [1]} as ever.

The Weyl reflections with respect to simple roots will be called
simple reflections $\sigma_i$. We extend multiple products of simple
reflections trivially by
$$ \sigma_{i_1.i_2}(\lambda) \equiv \sigma_{i_1}(\sigma_{i_2}(\lambda))  $$
and all that.

The length $\ell(\Sigma)$ of a Weyl group element $\Sigma$ with the
following reduced form
$$ \Sigma = \sigma_{i_1}.\sigma_{i_2} \dots \sigma_{i_k}
\eqno(I.1)$$ is known as $\ell(\Sigma) = k $ and its signature is
defined by
$$ \epsilon(\Sigma) \equiv (-1)^{\ell(\Sigma)}  . \eqno(I.2) $$
The reduced here means k is the minimum integer for $\Sigma$ .

In this work, we appropriately extend the concept of the signature
of a Weyl reflection to the concept of the signature of a weight. To
this end, let us consider that
$$ \mu = \sigma_{i_1.i_2 \dots i_s}(\lambda^+) \eqno(I.3)  $$
where $\mu$ is an element of the Weyl orbit $W(\lambda^+)$ of a
dominant weight $\lambda^+$. Our point of view here is to get (I.2)
in the absence of the knowledge of (I.3) and it will be given in the
next section.

We show in a subsequent paper {\bf [2]} that how this will be of
great help especially in applications of Weyl character formula for
higher Lie algebras like $E_8$.

\vskip 3mm \noindent {\bf{II.\ THE SIGNATURE OF A WEIGHT} } \vskip
3mm

As in sec.I, for any $\mu \in W(\lambda^+)$, we know that
$$ \lambda^+ - \mu \equiv \gamma^+ \in R^+(G_N) \eqno(II.1) $$
where $R^+(G_N)$ is the positive root lattice of $G_N$.

\eject

Our observation here is the following statement:

\noindent Let $ \Phi^+(G_N)$ be the positive root system of $G_N$
and \noindent $ \| \Phi^+(G_N) \| $ be its order  . Then,

\noindent (1) For $\beta_A \in \Phi^+(G_N)$, the equation $ \gamma^+
= \sum_{A=1}^{ \| \Phi^+(G_N) \|} \epsilon_A \ \beta_A $  has a
unique solution on condition that $ \epsilon_A=0,1 $ ,

\noindent (2) the number of $\epsilon_A$'s with the value +1 is
equal to the length of $\mu$, as defined in (I.3) and (I.2) .

For the validity of this statement, we only need the following observation.
For two positive integers N and M, let us consider
$$ \sigma_{i_1,i_2, \dots ,i_{N-1}}(\lambda^+) -
\sigma_{i_1,i_2, \dots ,i_N}(\lambda^+) \equiv \phi_N \in \Phi^+(G_N)  $$
and
$$ \sigma_{i_1,i_2, \dots ,i_{M-1}}(\lambda^+) -
\sigma_{i_1,i_2, \dots ,i_M}(\lambda^+) \equiv \phi_M \in \Phi^+(G_N) . $$
Then,for  $ N \neq M $, we know that $ \phi_N \neq \phi_M . $

Though this simple observation is sufficient to show the validity of
our statement above, it is also in the root  of the fact that the
order of Poincare series of finite Lie algebras is equal to the
order of their positive root systems. Note here that, the very
definition of Poincare series of a finite Lie algebra tells us that
its term of order k is equal to number of Weyl group elements of
length k.

One now is readily see that the signature of weight $\mu$ is defined as in the
following:
$$ \epsilon(\mu) \equiv (-1)^{k(\mu)}  $$
where
$$  k(\mu) \equiv \sum_{A=1}^{ \| \Phi^+(G_N) \|} \epsilon_A  \eqno(II.3)  $$

Though, this simple result helps for our subsequent works concerning
applications of Weyl character formula for $E_7$ and $E_8$ Lie
algebras, we would like to end up by giving here a nice example for
an element with length 29 of $E_8$ Weyl group.

\eject

\noindent Let us define $E_8$ as being in line with the
following Dynkin diagram:

\midinsert \epsfxsize=7cm \centerline{\epsfbox{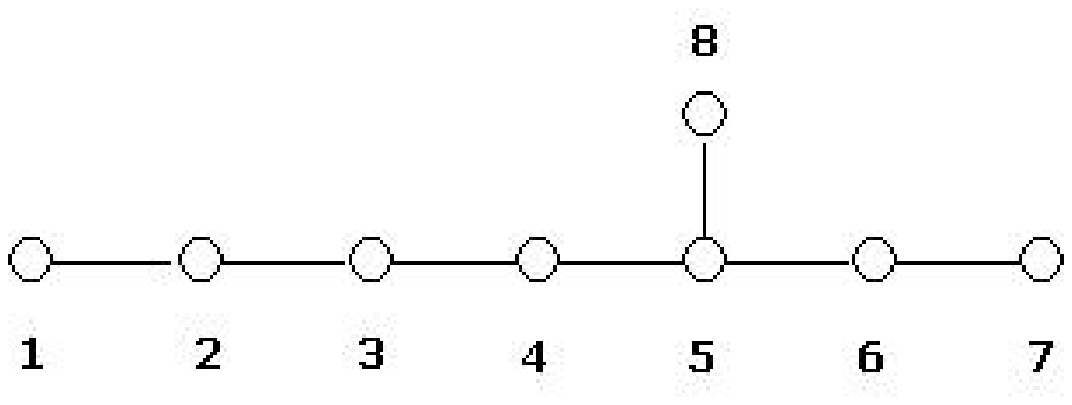}}
\endinsert

Let us consider an element $ \mu = \Sigma(\rho)  $ where
$$ \Sigma \equiv \sigma_{4,3,2,5,4,3,7,8,5,4,3,2,1,6,5,4,3,2,8,5,4,3,6,5,4,7,6,5,8}   \eqno(II.4) $$
with  $\ell(\Sigma)= 29 .$ To reach the same result but in another
way, one sees that
$$ \eqalign{
& \rho - \mu = \alpha_{114} + \alpha_{104} + \alpha_ {92} + \alpha_
{88} + \alpha_ {75} + \alpha_ {54} + \alpha_ {48} + \alpha_ {70} +
\alpha_ {49} + \cr & \alpha_ {42} + \alpha_ {65}  + \alpha_ {64} +
\alpha_ {43} + \alpha_ {36} + \alpha_ {60} + \alpha_ {55}  + \alpha_
{39} + \alpha_ {15} + \alpha_8 + \cr &\alpha_ {33} + \alpha_ {27} +
\alpha_ {21} + \alpha_7 + \alpha_ {24} + \alpha_ {18} + \alpha_ {12}
+ \alpha_ {17} + \alpha_ {11} + \alpha_4 } \eqno(II.6) $$ \noindent
where $\alpha_I \in \Phi(E_8) \ (I = 1,2 , \dots ,120 )$ aand hence
$k(\mu) = 29$ in the notation of (II.3). Ordering of the roots
$\alpha_I$ in (II.6) is as in the reduction of $\rho$ down to $\mu$
step by step. For example,
$$ \sigma_{4,3,2,5,4,3,7,8,5,4,3,2,1,6,5,4,3,2}(\rho) -
\sigma_{4,3,2,5,4,3,7,8,5,4,3,2,1,6,5,4,3,2,8}(\rho) = \alpha_{65} $$
or
$$ \sigma_{4,3,2,5,4,3,7,8}(\rho) -
\sigma_{4,3,2,5,4,3,7,8,5}(\rho) = \alpha_{27} $$ and all that. The
composite roots in (II.6) are given in appendix.

As the final note, let us emphasize that (II.5) is valid for $
\lambda^{++} - \mu^{++} $ where $ \lambda^{++} = \rho + \lambda^+ $
is a strictly dominant weight and $ \mu^{++} = \Sigma(\lambda^{++})
$ where $\Sigma$ is given in (II.4). Hence we conclude that the
signature of a weight is exactly the same with the signature of a
Weyl group element though the former is possible for actual
calculations.

\vskip3mm \noindent{\bf {REFERENCES}}

\item [1] Humphreys J.E., Introduction to Lie Algebras and Representation
\item \ \ \ \ Theory,  Springer-Verlag, 1972
\item [2] Karadayi H.R. and Gungormez M., Explicit Calculations of $E_8$
Characters
\item \ \ \ \ paper in preparation

\eject

\noindent{\bf {APPENDIX}

$$ \eqalign{
\alpha_{114} &= \alpha_1+2 \ \alpha_2+3 \ \alpha_3+4 \ \alpha_4+5 \ \alpha_5+3 \ \alpha_6+
2 \ \alpha_7+ 3 \ \alpha_8  \cr
\alpha_{104} &= \alpha_1 + 2 \ \alpha_2 + 3 \ \alpha_3 + 4 \ \alpha_4 + 4 \ \alpha_5 +
2 \ \alpha_6 + \alpha_7 + 2 \ \alpha_8   \cr
\alpha_{92} &= \alpha_2+2 \ \alpha_3+3 \ \alpha_4 + 4 \ \alpha_5+2 \ \alpha_6 + \alpha_7+2 \ \alpha_8 \cr
\alpha_{88} &= \alpha_2+2 \ \alpha_3+3 \ \alpha_4+3 \ \alpha_5+2 \ \alpha_6+ \alpha_7+2 \ \alpha_8 \cr
\alpha_{75} &= \alpha_1+2 \ \alpha_2+2 \ \alpha_3+2 \ \alpha_4+2 \ \alpha_5+ \alpha_6+\alpha_7+ \alpha_8 \cr
\alpha_{54}&=\alpha_2+ \alpha_3+ \alpha_4+2 \ \alpha_5+ \alpha_6+ \alpha_7+ \alpha_8 \cr
\alpha_{48}&=\alpha_2+ \alpha_3+ \alpha_4+ \alpha_5+ \alpha_6+ \alpha_7+ \alpha_8 \cr
\alpha_{70}&=\alpha_1+ \alpha_2+2 \ \alpha_3+2 \ \alpha_4+2 \ \alpha_5+ \alpha_6+ \alpha_7+ \alpha_8   \cr
\alpha_{49}&=\alpha_3+ \alpha_4+2 \ \alpha_5+ \alpha_6+ \alpha_7+ \alpha_8 \cr
\alpha_{42}&=\alpha_3+ \alpha_4+ \alpha_5+ \alpha_6+ \alpha_7+ \alpha_8 \cr
\alpha_{65}&=\alpha_2+2 \ \alpha_3+2 \ \alpha_4+2 \ \alpha_5+ \alpha_6+ \alpha_7+ \alpha_8 \cr
\alpha_{64}&=\alpha_1+ \alpha_2+ \alpha_3+2 \ \alpha_4+2 \ \alpha_5+ \alpha_6+ \alpha_7+ \alpha_8 \cr \alpha_{43}&=\alpha_4+2 \ \alpha_5+ \alpha_6+ \alpha_7+ \alpha_8 \cr
\alpha_{36}&=\alpha_4+ \alpha_5+ \alpha_6+ \alpha_7+ \alpha_8 \cr
\alpha_{60}&=\alpha_2+ \alpha_3+2 \ \alpha_4+2 \ \alpha_5+ \alpha_6+ \alpha_7+ \alpha_8 \cr
\alpha_{55}&=\alpha_3+2 \ \alpha_4+2 \ \alpha_5+ \alpha_6+ \alpha_7+ \alpha_8 \cr
\alpha_{39}&=\alpha_1+ \alpha_2+ \alpha_3+ \alpha_4+ \alpha_5+ \alpha_8 \cr
\alpha_{15}&=\alpha_5+ \alpha_8 \cr
\alpha_{33}&=\alpha_2+ \alpha_3+ \alpha_4+ \alpha_5+ \alpha_8 \cr
\alpha_{27}&=\alpha_3+ \alpha_4+ \alpha_5+ \alpha_8 \cr
\alpha_{21}&=\alpha_4+ \alpha_5+ \alpha_8 \cr
\alpha_{24}&=\alpha_2+ \alpha_3+ \alpha_4+ \alpha_5 \cr
\alpha_{18}&=\alpha_3+ \alpha_4+ \alpha_5 \cr
\alpha_{12}&=\alpha_4+ \alpha_5 \cr
\alpha_{17}&=\alpha_2+ \alpha_3+ \alpha_4 \cr
\alpha_{11}&=\alpha_3+ \alpha_4 }  $$

\end